\documentclass[12pt]{iopart}
\usepackage{graphicx}
\begin{document}
\title[Probing flavor changing neutral currents...]{Probing flavor changing neutral currents and CP violation through the decay $h\rightarrow c\overline{b} W^-$ in the two Higgs doublet model}
\author{L. D\'iaz-Cruz$^1$, A. D\'iaz-Furlong$^1$, R. Gait\'an-Lozano$^2$ and J.H. Montes de Oca Y.$^2$\footnote{Corresponding author: josehalim@comunidad.unam.mx}}

\address{$^1$Cuerpo Acad\'emico de Part\'iculas, Campos y Relatividad Facultad de Ciencias
F\'{\i}sico-Matem\'aticas, BUAP. Apdo. Postal 1364, C.P. 72000
Puebla, Pue., M\'exico}
\address{$^2$Departamento de F\'isica, Facultad de Estudios Superiores
Cuautitl\'an, UNAM, C.P. 54770, Estado de M\'exico, M\'exico}
\ead{jldiaz@fcfm.buap.mx}
\ead{rgaitan@servidor.unam.mx}
\ead{josehalim@comunidad.unam.mx}

\begin{abstract}
We discuss the formulation of the general two-Higgs doublet model
type III, which incorporates flavor changing neutral scalar
interactions (FCNSI) and CP violation from several sources. CP
violation can arise either from Yukawa terms or from the Higgs
potential, and it can explicit or spontaneous. We discuss the case that
includes CP violation with Yukawa textures to control FCNSI and
evaluate the CP asymmetry for the decay $h \rightarrow bcW$, which
may allow to test the patterns of FCNSI and CP violation, that
arises in these models.
\end{abstract}
\vspace{2pc}
\maketitle

\section{Introduction}
After many years of expectations, LHC is currently probing the mechanism of electroweak symmetry breaking (EWSB)\cite{Gunion:1989we}. In fact, LHC has provided great results and we know now that only an small mass window remain open for the SM Higgs mass, namely $116<m_{higgs}<130$ GeV \cite{atlas}. This  results are consistent with the analysis of electroweak precision of a Higgs mass around 125 GeV \cite{Erler:2010wa}. If this result is confirmed by future analysis, we will have one of the greatest discoveries of mankind. On the other hand, the SM is often considered as an effective theory, valid up to an energy scale of $O(TeV)$, that eventually will be replaced
by a more fundamental theory \cite{Bustamante:2009us}, which will explain, among other things, the physics behind EWSB and perhaps
even the origin of flavor. Many examples of candidate theories, which range from supersymmetry \cite{Ellis:2010wx} to strongly interacting models \cite{ArkaniHamed:2001nc} as well as some extra dimensional scenarios \cite{Aranda:2002dz}, include a multi-scalar Higgs sector. In particular, models with two scalar doublets have been studied extensively \cite{Haber:1978jt}, as they include a rich structure with interesting phenomenology
\cite{Carena:2000yx}.

Several versions of the 2HDM have been studied in the literature \cite{Accomando:2006ga}. Some models (known as 2HDM-I and 2HDM-II)
involve natural flavor conservation \cite{Glashow:1976nt}, while other models (known as 2HDM-III)  \cite{Accomando:2006ga}, allow for
the presence of flavor changing scalar interactions (FCNSI) at a level consistent with low-energy constraints \cite{DiazCruz:2004tr}.
There are also some variants (known as top, lepton, neutrino), where one Higgs doublet couples predominantly to one type of fermion
\cite{Atwood:2005bf}, while in other models it is even possible to identify a candidate for dark matter \cite{DiazCruz:2007be}. The definition of all these models, depends on the Yukawa structure and symmetries of the Higgs sector \cite{DiazCruz:2004ss,Aranda:2005st,DiazCruz:2006ki,Barbieri:2005kf,Froggatt:2007qp}, whose origin is still not known. The possible appearance of new sources of CP violation is another characteristic of these models \cite{Ginzburg:2005yw}.

In this paper we aim to discuss FCNC within general version of the Two-Higgs doublet model (2HDM-III), which incorporates flavor and CP
violation from all possible sources \cite{Maniatis:2007vn,Gerard:2007kn,ElKaffas:2007rq}. Within model I (2HDM-I) where only one Higgs doublet generates all gauge and fermion masses, while the second doublet only knows about this through mixing, and thus the Higgs phenomenology will share some similarities with the SM, although the SM Higgs couplings will now be shared among the neutral scalar spectrum. The presence of a charged Higgs boson is clearly the signal beyond the SM. Within 2DHM-II one also has natural flavor conservation \cite{Glashow:1976nt}, and its phenomenology will be similar to the 2HDM-I, although in this case the SM couplings are shared not only because of mixing, but also because of the Yukawa structure. The distinctive characteristic of 2HDM-III is the presence of FCNSI, which require a certain mechanism in order to suppress them, for instance one can imposes a certain texture for the Yukawa couplings \cite{Fritzsch:1977za}, which will then predict a pattern of FCNSI Higgs couplings \cite{Cheng:1987rs}. Within all those models (2HDM I,II,III) \cite{Carcamo:2006dp,Zhou:2003kd,Aoki:2009ha}, the Higgs doublets couple, in principle, with all fermion families, with a strength proportional to the fermion masses, modulo other parameters.

There are also other models where the Higgs doublets couple non-universally to the fermion families, which have also been
discussed in the literature \cite{Atwood:2005bf,Logan:2010ag,Logan:2009uf}. In principle, the general model includes CPV, which could arise from the same CPV phase that appears in the CKM matrix, as in the SM, from some other extra phase coming from the Yukawa sector or from the Higgs potential \cite{Gunion:2005ja}. However, in order to discuss which type of CP violation can appear in each case besides containing a generic pattern of FCNSI, moduled by certain texture, will include new sources of CPV as well.

In this paper we are interested in finding a signal of both FCNC and CPV. In particular, we identify the decay $h\rightarrow c\bar{b}W^-$ as a possible reaction where such test can be done. This decay process though the flavor current verex $tch$, which is not severely constrained by data, and is also sensitive to the presence of CPV phases. We define a decay asymmetry to probe the CPV. The organization of the paper goes as follows: section 2 includes the model formalism, which is based on the results of ref. \cite{Haber:2006ue}, and two considered scenarios. Section 3 contains the estimation of the width decay, branching ratio and CP asymmetry coefficient. Finally, the conclusions.
\section{General two Higgs doublet model with textures in Yukawa matrices}
The scalar field content of the model is two doublets under
$SU(2)_L$ with hypercharges $Y_1=Y_2=1/2$. The model is classified
by choice of the Higgs potential and the scalar-fermion couplings.
The simpler versions are considered when the potential is invariant
under $Z_2$ discrete symmetry and none of the doublets couples
simultaneously to the up-type and down-type fermions. These are
known as 2HDM type I and II. In type I, only one of the doublets
couples to fermions [gordon fest]. In type II, one doublet couples
to up-type fermions, the other to down-type fermions to prevent
tree-level FCNC's [gordon fest]. Type III is more general and
realistic model which allows all possible Higgs-fermion couplings
[gordon-sher]. This work is based on 2HDM type III with a general
potential which contains explicit and spontaneous CP violation.
\subsection{General Higgs potential and Higgs mass-eigenstates}
We follow closely the formalism developed and notation introduced by
\cite{Haber:2006ue}. The most general gauge invariant renormalizable Higgs scalar
potential in a covariant form with respect to global $U(2)$
transformation is given by
\begin{equation}
V=Y_{a,\overline{b}}\Phi_{\overline{a}}^\dag\Phi_b+\frac{1}{2}Z_{a\overline{b}c\overline{d}}\left(\Phi_{\overline{a}}^\dag\Phi_b\right)
\left(\Phi_{\overline{c}}^\dag\Phi_d\right),
\end{equation}
where $\Phi_a=\left(\phi_a^+,\,\phi_a^0 \right)^T$ and $a,b,c,d$ are
labels with respect to two dimensional Higgs flavor space. The index
conventions means that replacing an unbarred index with a barred
index is equivalent to complex conjugation and barred.unbarred index
pair denotes a sum. The most general $U(1)_{EM}$-conserving vacuum
expectation values are
\begin{equation}
\langle \Phi_a \rangle= \frac{v}{\sqrt{2}}\left(
\begin{array}{c}
0 \\
\hat{v}_a \\
\end{array}
\right),
\end{equation}
where $\left(\begin{array}{cc}\hat{v}_1, & \hat{v}_2 \\ \end{array}
\right) =e^{i\eta}\left(\begin{array}{cc} \cos \beta,& \sin\beta
e^{i\xi}\\ \end{array}\right)$ and $v=246$ GeV. The
covariant form for the scalar potential minimum conditions is
\begin{equation}
v\hat{v}_{\bar{a}}^*\left[
Y_{a\bar{b}}+\frac{1}{2}v^2Z_{a\bar{b}c\bar{d}}
\hat{v}_{\bar{c}}^*\hat{v}_{d}\right]=0.
\end{equation}
The Higgs mass-eigenstates of the neutral Higgs bosons are
explicitly derived in Ref. \cite{Haber:2006ue}. The expressions for Higgs bosons in
terms of the generic basis is
\begin{eqnarray}
h_k&=&\frac{1}{\sqrt{2}}\overline{\Phi}_{\overline{a}}^{0\dag}(q_{k1}\hat{v}_a+q_{k2}\hat{w}_ae^{-i\theta_{23}})\nonumber\\
&&\frac{1}{\sqrt{2}}(q_{k1}^*\hat{v}_{\overline{a}}^*+q_{k2}*\hat{w}_{\overline{a}}^*e^{i\theta_{23}})\overline{\Phi}_{\overline{a}}^{0}
\label{h-1}
\end{eqnarray}
for $k=1,...,4$, where $h_{1,2,3}$ are the neutral physical Higgs
bosons and $h_4=G^0$ is the goldstone boson. The parameters
$q_{k1,2}$ are functions of the neutral Higgs mixing angles and the
explicit values are shown in table \ref{table-q}.
\begin{table}
\begin{center}
\begin{tabular}{|c|c|c|}
\hline $r$ & $q_{r1}$ & $q_{r2}$ \\ \hline
1 & $\cos \theta_{12}\cos \theta _{13}$ & $-\sin \theta _{12}-i\cos \theta_{12}\sin\theta_{13}$\\
2 & $\sin \theta _{12}\cos \theta _{13}$ & $\cos \theta _{12}-i\sin \theta_{12}\sin\theta_{13}$ \\
3 & $\sin \theta _{13}$ & $i\cos \theta _{13}$ \\
\hline
\end{tabular}
\caption{Mixing angles for Higgs bosons which consider spontaneous
and explicit CPV \cite{Haber:2006ue}.} \label{table-q}
\end{center}
\end{table}
It is possible to invert the expression (\ref{h-1}) and the result
is given by
\begin{equation}
\Phi_a=\left(
         \begin{array}{c}
           G^+\hat{v}_a+H^+\hat{w}_a \\
           \frac{v}{\sqrt{2}}\hat{v}_a+\frac{1}{\sqrt{2}}\sum_{k=1}^{4}\left(q_{k1}\hat{v}_a+q_{k2}e^{-\theta_{23}}\hat{w}_a\right)h_k \\
         \end{array}
       \right),
       \label{phis}
\end{equation}
where $\hat{w}_a^T=e^{-i\eta}\left(\begin{array}{cc} -\sin \beta
e^{-i\xi},& \cos\beta\\ \end{array}\right)$.
\subsection{Higgs-fermions couplings}
We focus on quarks fields with analogous treatment for leptons. The
most general structure of the Yukawa lagrangian for the quark
fields can be written as follows:
\begin{equation}
\mathcal{L}_{Y}^{quarks}=\overline{q}_{L}^{0}Y_{1}^{D}\Phi _{1}d_{R}^{0}+%
\overline{q}_{L}^{0}Y_{2}^{D}\Phi _{2}d_{R}^{0}+\overline{q}_{L}^{0}Y_{1}^{U}%
\widetilde{\Phi }_{1}u_{R}^{0}+\overline{q}_{L}^{0}Y_{2}^{U}\widetilde{\Phi }%
_{2}u_{R}^{0}+h.c.,  \label{yukawa}
\end{equation}
where $Y_{1,2}^{U,D}$ are the $3\times 3$ Yukawa matrices, $q_{L}$
denotes the left handed quarks doublets and $u_{R}$, $d_{R}$
correspond to the right handed singlets. The superscript zero means
that the quarks are weak eigenstates. Here $\widetilde{\phi
}_{1,2}=i\sigma _{2}\phi _{1,2}^{\ast }$. After getting a correct SSB
\cite{PortugalMinHix,Ivanov:2006yq,Maniatis:2006fs, Ma:2010ya}, the
Higgs doublets are decomposed as shown in (\ref{phis}) and the
neutral scalar and pseudoscalar couplings within up-type quarks in
mass eigenstate basis are
\begin{equation}
\mathcal{L}_{up}^{neutral}=\overline{u}_{i}\left(S_{ijr}^{u}+%
\gamma^{5}P_{ijr}^{u}\right)
u_{j}H_{r}+\overline{u}_{i}M_{ij}^{U}u_{j}, \label{ygral}
\end{equation}
where we have denoted the scalar and pseudoscalar couplings as
\begin{eqnarray}
S_{ijr}^{u} &=&\frac{1}{v}M_{ij}^{U}\left( q_{r1}+\cot \beta
\textrm{Re} \left[q_{r2}e^{-i\left(\theta _{23}+\xi \right) }\right] \right) \nonumber \\
&&+\frac{1}{2\sqrt{2}\sin \beta }\left( q_{r2}^{\ast }e^{i\theta_{23}}%
\widetilde{Y}_{2ij}^{U}+q_{r2}e^{-i\theta _{23}}\widetilde{Y}%
_{2ij}^{U\dagger }\right)   \label{sugral}
\end{eqnarray}
and
\begin{eqnarray}
P_{ijr}^{u} &=&\frac{1}{v}M_{ij}^{U}\cot \beta \textrm{Im}
\left[q_{r2}e^{-i\left(\theta _{23}+\xi \right) }\right] \nonumber \\
&&+\frac{1}{2\sqrt{2}\sin \beta }\left( q_{r2}^{\ast }e^{i\theta _{23}}%
\widetilde{Y}_{2ij}^{U}-q_{r2}e^{-i\theta _{23}}\widetilde{Y}%
_{2ij}^{U\dagger }\right),   \label{pugral}
\end{eqnarray}
respectively. The mass matrices are given as follows:
\begin{equation}
M^{U}=\frac{v_{1}}{\sqrt{2}}\widetilde{Y}_{1}^{U}+e^{-i\xi }\frac{v_{2}}{%
\sqrt{2}}\widetilde{Y}_{2}^{U}  \label{mu}
\end{equation}%
and
\begin{equation}
M^{D}=\frac{v_{1}}{\sqrt{2}}\widetilde{Y}_{1}^{D}+e^{i\xi }\frac{v_{2}}{%
\sqrt{2}}\widetilde{Y}_{2}^{D},  \label{md}
\end{equation}
where $\widetilde{Y}_{1,2}^U=U_LY_{1,2}^UU_R^\dagger$ and
$\widetilde{Y}_{1,2}^D=D_LY_{1,2}^D D_R^\dagger$ with
$u_{L,R}=U_{L,R}u_{L,R}^{0}$ and $d_{L,R}=D_{L,R}d_{L,R}^{0}$. The
vacuum expectation values $v_1$ and $v_2$ are real and positive, while the phase $\xi$
introduces spontaneous CP violation. Analogously, the down-type quarks are
\begin{equation}
\mathcal{L}_{down}^{neutral}=\overline{d}_{i}\left(
S_{ijr}^{d}+\gamma ^{5}P_{ijr}^{d}\right)
d_{j}H_{r}+\overline{d}_{i}M_{ij}^{D}d_{j},
\end{equation}
with
\begin{eqnarray}
S_{ijr}^{d} &=&\frac{1}{v}M_{ij}^{D}\left( q_{r1}-\tan \beta
\textrm{Re} \left[q_{r2}e^{-i\left(\theta _{23}+\xi \right) }\right] \right)   \nonumber \\
&&+\frac{1}{2\sqrt{2}\cos \beta }\left( q_{r2}e^{-i\theta
_{23}}Y_{2}^{D}+q_{r2}^{\ast }e^{i\theta
_{23}}\widetilde{Y}_{2}^{D\dagger }\right)   \label{sdgral}
\end{eqnarray}
and
\begin{eqnarray}
P_{ijr}^{d} &=&-\frac{1}{v}M_{ij}^{D}\tan \beta
\textrm{Im}\left[q_{r2}e^{-i\left(\theta _{23}+\xi \right) }\right]   \nonumber \\
&&+\frac{1}{2\sqrt{2}\cos \beta }\left( q_{r2}e^{-i\theta _{23}}\widetilde{Y}%
_{2}^{D}-q_{r2}^{\ast }e^{i\theta _{23}}\widetilde{Y}_{2}^{D\dagger
}\right). \label{pdgral}
\end{eqnarray}
For completeness, the Yukawa couplings for charged Higgs
bosons is written as
\begin{eqnarray}
\mathcal{L}_{Y}^{H^+} &=&\overline{u}\left[ H^{+}e^{-i\xi
}M^{U}V\frac{I -\gamma ^{5}}{\sqrt{2}}-H^{+}e^{-i\xi
}VM^{D}\frac{I+\gamma ^{5}}{\sqrt{2}}\right.   \nonumber \\
&&\left. +\frac{1}{\cos \beta }H^{+}\left( V\widetilde{Y}_{2}^{D}\frac{%
I+\gamma ^{5}}{2}-\widetilde{Y}_{2}^{U\dag }V\frac{I%
-\gamma ^{5}}{2}\right) \right] d  \nonumber \\
&&+h.c.,  \label{lch}
\end{eqnarray}
where $V$ denotes the CKM matrix and physical eigenstates for the
charged Higgs boson $(H^\pm)$ can be obtain through (\ref{phis}).
\subsection{Universal Yukawa textures}
Suppression for FCNC can be achieved when a certain form of the
Yukawa matrices that reproduce the observed fermion masses and
mixing angles is implemented in the model. This could be done either
by implementing the Frogart-Nielsen mechanism to generate the
fermion mass hierarchies \cite{FN}, or by studying a certain ansatz
for the fermion mass matrices \cite{Fritzsch:1977za}. The first
proposal for the Higgs boson couplings \cite{Cheng:1987rs}, the so
called Cheng-Sher ansazt, was based on the Fritzsch six-texture form
of the mass matrices, namely:
\begin{equation}
M_{l}=\left(
\begin{array}{ccc}
0 & C_{q} & 0 \\
C_{q}^{\ast } & 0 & B_{q} \\
0 & B_{q}^{\ast } & A_{q}%
\end{array}%
\right) .
\end{equation}%
Then, by assuming that each Yukawa matrix $Y_{1,2}^{q}$ has the same
hierarchy, one finds: $A_{q}\simeq m_{q_{3}}$, $B_{q}\simeq \sqrt{%
m_{q_{2}}m_{q_{3}}}$ and $C_{q}\simeq \sqrt{m_{q_{1}}m_{q_{2}}}$.
Then, the fermion-fermion$^{\prime }$-Higgs boson ($f f^{\prime 0}$)
couplings obey the following pattern: $Hf_{i}f_{j} \sim
\sqrt{m_{f_i}m_{f_j}} / m_{W}$, which is also known as the
Cheng-Sher ansatz. This brings under control the FCNC problem, and
it has been extensively studied in the literature to search for
flavor-violating signals in the Higgs sector 

In our previous work we considered in detail the case of universal
four-texture Yukawa matrices \cite{DiazCruz:2004tr}, and derived the
scalar-fermion interactions, showing that it was possible to satisfy
current constraints from LFV and FCNC \cite{ourwork1,ourwork2}.
Predictions for Higgs phenomenology at the LHC was also studied in
ref. \cite{ourwork3,otherswork1}. We can consider this  a universal
model, in the sense that it was assumed that each Yukawa matrix
$Y^q_{1,2}$ has the same hierarchy.
\subsection{FCNC and CPV Feynman rules}
The Higgs-fermions interactions and the scalar potential of the general 2HDM
contain several sources of CPV and FCNC. In order to explore these
sources we consider some limiting cases. As it is discussed in
previous sections, the assumption of universal 4-textures for the
Yukawa matrices, allows to express one Yukawa matrix in terms of the
quark masses, and parametrization of the FCNSI in terms of the unknown
coefficients $\chi _{ij}$, namely $\widetilde{Y}_{2ij}^{U}=\chi
_{ij}\frac{\sqrt{ m_{i}m_{j}}}{v}$. These parameters can be
constrained by considering all types of low energy FCNC transitions.
Although these constraints are quite strong for transitions
involving the first and second families, as well as for the b-quark,
it turns out that they are rather mild for the top quark. In this paper we shall consider two scenarios:\\
\emph{A) Explicit CPV in the Higgs sector}. In this case we assume the hermiticity condition for the Yukawa
matrices, but the Higgs sector admits explicit CP violating. Then, from (\ref{sugral}) and (\ref{pugral}), one obtains the following expressions for the couplings of the neutral Higgs bosons with up-type quarks, namely:
\begin{equation}
S_{ijr}^{u}=\frac{1}{v}M_{ij}^{U}\left(q_{r1}+\cot \beta
\textrm{Re}\left[ q_{r2}\right] \right)
+\frac{\sqrt{m_{i}m_{j}}}{\sqrt{2}v\sin \beta }\chi _{ij}\left(
\textrm{Re}\left[q_{r2}\right]\right) \label{sva}
\end{equation}
and
\begin{equation}
P_{ijr}^{u}=\frac{i}{v}M_{ij}^{U} \cot \beta \textrm{Im}\left[
q_{r2}\right], \label{pva}
\end{equation}
similar expressions can be obtained for the down-type quarks and leptons.\\
\emph{B) CP conserving Higgs sector}. In this case we shall consider that the Higgs sector is CP
conserving and the Yukawa matrices could be also hermitian. Then, without loss of generality, we can assume that $h_{3}$ is CP odd, while $h_{1}$ and $h_{2}$ are CP even. Then: $\cos \theta _{12}=\sin \left( \beta -\alpha \right)$, $\sin \theta _{12}=\cos \left( \beta
-\alpha \right)$, $\sin \theta _{13}=0$, and $ e^{-i\theta_{13}}=1$. The mixing angles $\alpha $ and $\beta $ that appear in
the neutral Higgs mixing, corresponds to the standard notation. Additionally, when one assumes a 4-texture for the Yukawa matrices,
the Higgs-fermion couplings further simplify as $\widetilde{Y}_{2ij}^{U}=\chi _{ij}\frac{\sqrt{m_{i}m_{j}}}{v}$. Then, the corresponding coefficients for up sector and $h^0$ ($r=1$) are
\begin{equation}
S_{ij1}^{u}=\frac{1}{v}M_{ij}^{U}\left[ \sin (\beta -\alpha )+\cot \beta
\cos (\beta -\alpha )\right] +\frac{\sqrt{m_{i}m_{j}}\chi _{ij}}{\sqrt{2}v}%
\frac{\cos (\beta -\alpha )}{\sin \beta }  \label{svb}
\end{equation}
For $H^0$ $(r=2)$ one finds:
\begin{equation}
S_{ij2}^{u}=\frac{1}{v}M_{ij}^{U}\left[ \cos (\beta -\alpha )+\cot \beta
\sin (\beta -\alpha )\right] +\frac{\sqrt{m_{i}m_{j}}\chi _{ij}}{\sqrt{2}v}%
\frac{\sin (\beta -\alpha )}{\sin \beta }.
\end{equation}
Finally, for $A^0$ $(r=3)$ one obtains:
\begin{equation}
P_{ij3}^{u}=\frac{i}{2v}M_{ij}^{U}\cot \beta -i\frac{\sqrt{m_{i}m_{j}}\chi
_{ij}}{2\sqrt{2}v\sin \beta }  \label{pvb3}
\end{equation}
Note that under hermiticity of the Yukawa matrices the $P_{ijr}^u$
couplings for the $h^0$ and $H^0$ and the coupling $S_{ij3}^u$ for the $A^0$ are
vanished.
\section{The Higgs decay $h\rightarrow c\overline{b} W^{-}$}
\begin{figure}
\begin{minipage}{14pc}
\includegraphics[width=18pc]{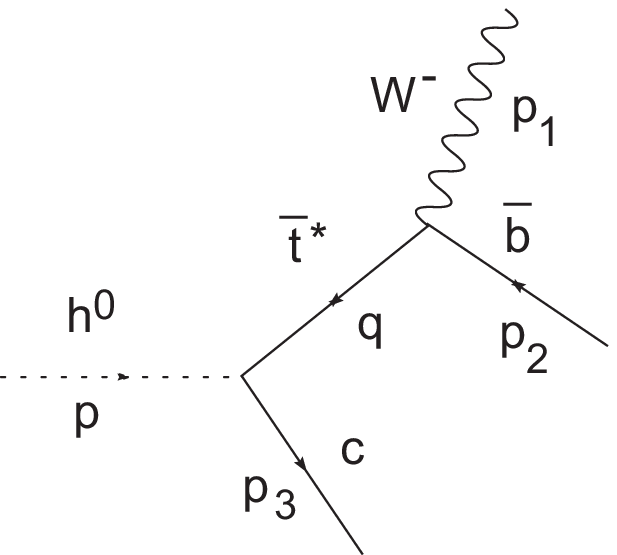}
\caption{\label{f1a}Tree level Feynman diagrams for
$h\longrightarrow W^{-}\overline{b}c$.}
\end{minipage}\hspace{3pc}%
\begin{minipage}{14pc}
\includegraphics[width=18pc]{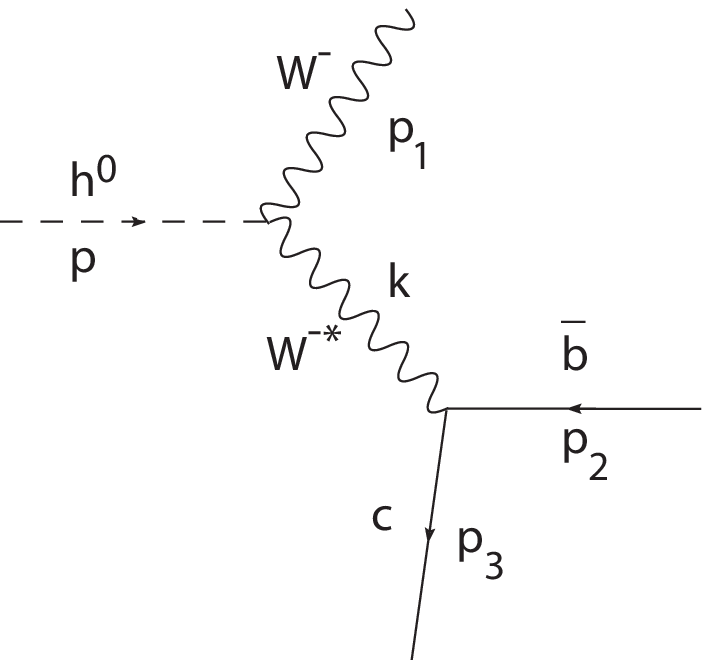}
\caption{\label{f1b}Tree level Feynman diagrams for
$h\longrightarrow W^{-}\overline{b}c$.}
\end{minipage}
\end{figure}

In this section we shall evaluate the width of the decay $h\rightarrow c\bar{b}W^-$ and its respective branching ratio. Besides the decay width and branching ratio, we are also interested in defining a decay asymmetry coefficient for the
decay $h\to c\bar{b}W$ in order to analyze presence of both FCNSI and CPV within the 2HDM. We consider the neutral Higgs boson decay $h\longrightarrow W\overline{b}c$\ at tree level. Two diagrams contribute to this decay, the first one is through the FCNC coupling $h\longrightarrow \overline{t}^{\ast }c\longrightarrow W^{-}\overline{b}c$, its Feynman diagram is shown on figure \ref{f1a}. The other one goes through $h\longrightarrow W^{+\ast }W^{-}\longrightarrow W^{-}\overline{b}c$, see figure \ref{f1b}. The amplitude for these diagrams is thus
\begin{equation}
\overline{\vert \mathcal{M}\vert }^{2}=\overline{\vert
\mathcal{M}_1\vert}^{2}+\overline{\vert
\mathcal{M}_2\vert}^{2}+\overline{ \mathcal{M}_{1}^{\dagger }\mathcal{M}_{2}}+\overline{%
\mathcal{M}_{2}^{\dagger}\mathcal{M}_{1}}
\end{equation}
We can obtain an approximation when the terms proportional to the
charm and bottom masses are neglected. Then, the explicit
expressions are
\begin{eqnarray}
\overline{\left\vert M_{1}\right\vert }^{2} &=&\frac{g^{4}}{4M_{W}^{2}}%
\left\vert P_{t}\right\vert ^{2}\left\{ \left\vert
S_{231}^{u}+P_{231}^{u}\right\vert ^{2}\left[ 4\left( p_{1}\cdot
p_{2}\right) \left( p_{1}\cdot q\right) \left( p_{3}\cdot q\right) \right.
\right.   \nonumber \\
&&\left. +2M_{W}^{2}\left( p_{2}\cdot q\right) \left( p_{3}\cdot q\right)
-M_{W}^{2}q^{2}\left( p_{2}\cdot p_{3}\right) -2q^{2}\left( p_{1}\cdot
p_{2}\right) \left( p_{1}\cdot p_{3}\right) \right]   \nonumber \\
&&\left. +m_{t}^{2}\left\vert S_{231}^{u}-P_{231}^{u}\right\vert ^{2}\left[
M_{W}^{2}\left( p_{2}\cdot p_{3}\right) +2\left( p_{1}\cdot p_{2}\right)
\left( p_{1}\cdot p_{3}\right) \right] \right\} ,
\end{eqnarray}
\begin{equation}
\overline{\left\vert \mathcal{M}_{2}\right\vert }^{2}=g^{4}\left(
q_{11}\right) ^{2}\left\vert V_{cb}\right\vert
^{2}|P_{W}(k)|^{2}\left( M_{W}^{2}p_{2}\cdot p_{3}+2p_{1}\cdot
p_{2}p_{1}\cdot p_{3}\right) ,  \label{m22f}
\end{equation}
\begin{eqnarray}
\overline{\mathcal{M}_{1}^{\dagger }\mathcal{M}_{2}}+\overline{\mathcal{M}%
_{2}^{\dagger }\mathcal{M}_{1}} &=&\frac{g^{4}m_{t}}{M_{W}}%
q_{11}V_{cb}P_{W}(k)\left( M_{W}^{2}p_{2}\cdot p_{3}+2p_{1}\cdot
p_{2}p_{1}\cdot p_{3}\right)   \nonumber \\
&&\left[ \textrm{Re}\left( S_{231}^{u}+P_{231}^{u}\right)
\textrm{Re}P_{t}-\textrm{Im}\left( S_{231}^{u}+P_{231}^{u}\right)
\textrm{Im}P_{t}\right], \label{m1m2f}
\end{eqnarray}
where $P_{W}(k)=\left( k^{2}-M_{W}^{2}+iM_{W}\Gamma
_{W}\right)^{-1}$ and $P_{t}\left( q\right) =\left(
q^{2}-m_{t}^{2}+im_{t}\Gamma _{t}\right) ^{-1}$. Then, the decay width is
\begin{eqnarray}
\Gamma _{h\rightarrow W\overline{b}c}&=&\frac{g^4m_{h}}{256\pi
^{3}}[\left|S_{231}^u-P_{231}^u
\right|^2I_1+\left|S_{231}^u+P_{231}^u\right|^2I_2+q_{11}^2\left|V_{cb}\right|^2I_3\nonumber\\
&&+\textrm{Re}(S_{231}^u+P_{231}^u)q_{11}V_{cb}I_4-\textrm{Im}(S_{231}^u+P_{231}^u)q_{11}V_{cb}I_5]
\end{eqnarray}
The analytical expressions and numerical values for the integrals $I_{1,...,5}$ are shown in appendix. For scenario \emph{A}
\begin{eqnarray}
\Gamma _{h\rightarrow W\overline{b}c}^{{\tiny {\textrm{A}}}} &=&\frac{%
g^{4}m_{h}}{256\pi ^{3}}\left\{ \left\vert \chi _{23}\right\vert
^{2}m_{c}m_{t}v^{-2}\frac{\sin ^{2}\theta _{12}}{2\sin ^{2}\beta }\left(
I_{1}+I_{2}\right) \right.   \nonumber \\
&&+\left\vert V_{cb}\right\vert ^{2}\cos ^{2}\theta _{12}\cos ^{2}\theta
_{13}I_{3}  \nonumber \\
&&+V_{cb}\sqrt{m_{c}m_{t}}v^{-1}\frac{\sin \theta _{12}\cos \theta _{12}\cos
\theta _{13}}{2\sin \beta }  \nonumber \\
&&\left. \left[ I_{5}\textrm{Im}\left( \chi _{23}\right) -I_{4}\textrm{Re}\left(
\chi _{23}\right) \right] \right\} ,
\end{eqnarray}%
while for scenario \emph{B} we have:
\begin{eqnarray}
\Gamma _{h\rightarrow W\overline{b}c}^{{\tiny {\textrm{B}}}} &=&\frac{%
g^{4}m_{h}}{256\pi ^{3}}\left[ m_{c}m_{t}v^{-2}\frac{\cos ^{2}\left( \beta
-\alpha \right) }{2\sin ^{2}\beta }\left\vert \chi _{23}\right\vert
^{2}\left( I_{1}+I_{2}\right) \right.   \nonumber \\
&&+\left\vert V_{cb}\right\vert ^{2}\sin ^{2}\left( \beta -\alpha \right)
I_{3}  \nonumber \\
&&-V_{cb}\sqrt{m_{c}m_{t}}v^{-1}\frac{\sin \left( \beta -\alpha \right) \cos
\left( \beta -\alpha \right) }{2\sin \beta }  \nonumber \\
&&\left. \left( I_{4}\textrm{Re}\left( \chi _{23}\right) -I_{5}\textrm{Im}\left(
\chi _{23}\right) \right) \right] ,
\end{eqnarray}
\subsection{Branching ratio for $h\rightarrow c\bar{b}W^-$}
Since the analysis of electroweak precision test favors a Higgs mass in the range $115<m_h<160$ GeV, we should consider Higgs decay modes that could complete with the dominant decay in this window, which in most models is $h\rightarrow b\bar{b}$, although $h\rightarrow WW^*$ $h\rightarrow ZZ^*$ could also significant contribution. Furthermore, given the recent LHC data, one could assume that the light Higgs boson will have SM like couplings, but current LHC precision admits some moderate deviations from SM. Thus, we shall include the 3-body decay $h\rightarrow c\bar{b}W^-$, which could receive an enhancement coming from the vertex $ht^*c$.
The total width for the littlest neutral Higgs boson can be approximated by
\begin{eqnarray}
\Gamma _{Total} &\approx&\Gamma (h\rightarrow b\bar{b})+\Gamma (h\rightarrow c\bar{%
c})+\Gamma (h\rightarrow ZZ^{\ast })  \nonumber \\
&&+\Gamma (h \rightarrow WW^{\ast })+\Gamma (h\rightarrow c\bar{b}W^{-}),
\label{wtotal}
\end{eqnarray}
The decay width for the decay of the Higgs boson to fermion and anti-fermion
pair at tree level is given by
\begin{eqnarray}
\Gamma \left( h\rightarrow f_{i}\overline{f_{j}}\right)  &=&\frac{g^{2}m_{h}%
}{8\pi }\left( 1-4\frac{m_{f}^{2}}{m_{h}^{2}}\right)   \nonumber \\
&&\left[ \left( \left\vert S_{ij1}^{f}+P_{ij1}^{f}\right\vert
^{2}+\left\vert S_{ij1}^{f}-P_{ij1}^{f}\right\vert ^{2}\right) \left( 1-2%
\frac{m_{f}^{2}}{m_{h}^{2}}\right) \right.   \nonumber \\
&&\left. -\textrm{Re}\left[ \left( S_{ij1}^{f}+P_{ij1}^{f}\right) \left(
S_{ij1}^{f\ast }-P_{ij1}^{f\ast }\right) \right] \right]
\end{eqnarray}
In general, the couplings $S^f_{ij1}$ and $P^f_{ij1}$ are given by the equations (\ref{sugral}) and (\ref{pugral}) for any of the proposed scenarios. Now, the decay width into a real $W$ and virtual $W^*$ is given by \cite{papaqui}
\begin{equation}
\Gamma \left( h\rightarrow WW^{\ast }\right) =\frac{g^{4}m_{h}}{128\pi ^{3}}%
\cos ^{2}\left( \theta_{12}\right) F\left( \frac{m_{W}}{m_{h}}\right) .
\end{equation}
While the decay width $h\rightarrow ZZ^{\ast }$\ is given by
\begin{eqnarray}
\Gamma \left( h\rightarrow ZZ^{\ast }\right)  &=&\frac{g^{4}m_{h}}{2048\pi
^{3}}F\left( \frac{m_{Z}}{m_{h}}\right) \cos ^{2}\left( \theta_{12}\right)   \nonumber \\
&&\frac{7-\frac{40}{3}\sin ^{2}\theta _{W}+\frac{160}{9}\sin ^{4}\theta _{W}%
}{\cos ^{4}\theta _{W}}
\end{eqnarray}
where%
\begin{eqnarray}
F\left( x\right)  &=&-\left( 1-x^{2}\right) \left( \frac{47}{2}x^{2}-\frac{13%
}{2}+\frac{1}{x^{2}}\right) -3\left( 1-6x^{2}+4x^{4}\right) \ln \left(
x\right)  \nonumber\\
&&+3\frac{1-8x^{2}+20x^{4}}{\sqrt{4x^{2}-1}}\cos ^{-1}\left( \frac{3x^{2}-1}{%
2x^{3}}\right) .
\end{eqnarray}
It is possible to write the explicit expressions of the total decay width for each scenario. However, the section \ref{secN} shows the behavior for the branching ratio and the CP asymmetry coefficient.
\subsection{CPV decay asymmetry}
In order to find the CP asymmetry coefficient we also need to
calculate the conjugate decay. We denote the average amplitude of
the conjugate decay as
\begin{equation}
\overline{\vert \widetilde{\mathcal{M}}\vert }^{2} =\overline{\vert \widetilde{\mathcal{M}_{1}}%
\vert}^{2} +\overline{\vert\widetilde{\mathcal{M}_{2}}\vert}^{2}+\overline{\widetilde{%
\mathcal{M}_{1}}^{\dagger}\widetilde{\mathcal{M}_{2}}} +\overline{\widetilde{\mathcal{M}_{2}}^{\dagger }%
\widetilde{\mathcal{M}_{1}}}.
\end{equation}
The square terms are the same as the above, $\overline{\left\vert \widetilde{%
\mathcal{M}}_{1,2}\right\vert }^{2}=\overline{\left\vert
\mathcal{M}_{1,2}\right\vert }^{2}$, while for the interference
terms we have
\begin{eqnarray}
\overline{\widetilde{\mathcal{M}}_{1}^{\dagger }\widetilde{\mathcal{M}}_{2}}+%
\overline{\widetilde{\mathcal{M}}_{2}^{\dagger }\widetilde{\mathcal{M}}_{1}}
&=&\frac{g^{4}m_{t}}{M_{W}}q_{11}V_{cb}P_{W}(k)\left( M_{W}^{2}p_{2}\cdot
p_{3}\right)   \nonumber \\
&&\left. +2p_{1}\cdot p_{2}p_{1}\cdot p_{3}\right) \left[ \textrm{Re}\left(
S_{231}^{u}+P_{231}^{u}\right) \textrm{Re}(P_{t})\right.   \nonumber \\
&&\left. +\textrm{Im}\left( S_{231}^{u}+P_{231}^{u}\right) \textrm{Im}(P_{t})%
\right] .
\end{eqnarray}
The CP asymmetry coefficient is defined as
\begin{equation}
A_{CP}=\frac{\Gamma _{h\rightarrow W^+\overline{b}c}-\Gamma
_{h\rightarrow W^-b\overline{c}}}{\Gamma _{h\rightarrow W^+\overline{b}%
c}+\Gamma _{h\rightarrow W^-b\overline{c}}}.  \label{asy}
\end{equation}
Therefore, the CP asymmetry coefficient of the decay in general 2HDM
is given by:
\begin{eqnarray}
A_{CP} &=&V_{cb}q_{11}I_{5}\textrm{Im}\left( S_{231}^{u}+P_{231}^{u}\right) %
\left[ \left\vert S_{231}^{u}-P_{231}^{u}\right\vert ^{2}I_{1}+\left\vert
S_{231}^{u}+P_{231}^{u}\right\vert ^{2}I_{2}\right.   \nonumber \\
&&\left. +\left\vert V_{cb}\right\vert ^{2}q_{11}^{2}I_{3}+q_{11}V_{cb}\textrm{Re}
\left( S_{231}^{u}+P_{231}^{u}\right) I_{4}\right] ^{-1},
\label{asymmetry1}
\end{eqnarray}
In case of CP explicit violation or CP conserving the asymmetry
coefficient is written as
\begin{equation}
A_{h\rightarrow W\overline{b}c}^{{\tiny {\textrm{A}}}}=\frac{\textrm{Im}%
\left( \chi _{23}\right) I_{5}}{f_{{\tiny {\textrm{A}}}}\left( \theta
_{12},\theta _{13},\beta ,\chi _{23}\right) }
\label{acpa}
\end{equation}
or
\begin{equation}
A_{h\rightarrow W\overline{b}c}^{{\tiny {\textrm{B}}}}=\frac{I_{5}\textrm{Im%
}\left( \chi _{23}\right) }{f_{{\tiny {\textrm{B}}}}\left( \alpha ,\beta
,\chi _{23},m_{h}\right) },
\label{acpb}
\end{equation}
respectively. The $f_{\tiny{\textrm{A}}}$ and $f_{\tiny{\textrm{B}}}$ are defined as
\begin{eqnarray}
f_{{\tiny {\textrm{A}}}} &=&I_{4}\textrm{Re}\left( \chi _{23}\right) -%
\sqrt{2}\frac{I_{3}V_{cb}v}{\sqrt{m_{c}m_{t}}}\frac{\cos \theta _{13}\sin
\beta }{\tan \theta _{12}}  \nonumber \\
&&-\left( I_{1}+I_{2}\right) \frac{\left\vert \chi _{23}\right\vert ^{2}%
\sqrt{m_{c}m_{t}}}{\sqrt{2}v}\frac{\tan \theta _{12}}{\sin \beta \cos \theta
_{13}}
\end{eqnarray}
and%
\begin{eqnarray}
f_{{\tiny {\textrm{B}}}} &=&I_{4}\textrm{Re}\left( \chi _{23}\right) -\sqrt{%
2}\frac{I_{3}V_{cb}v}{\sqrt{m_{c}m_{t}}}\sin \beta \tan \left( \beta -\alpha
\right)   \nonumber \\
&&-\left( I_{1}+I_{2}\right) \frac{\left\vert \chi _{23}\right\vert ^{2}%
\sqrt{m_{c}m_{t}}}{\sqrt{2}v\sin \beta \tan \left( \beta -\alpha \right) }
\end{eqnarray}
\subsection{Numerical results}
\label{secN}
\begin{figure}
\begin{minipage}{14pc}
\includegraphics[width=18pc]{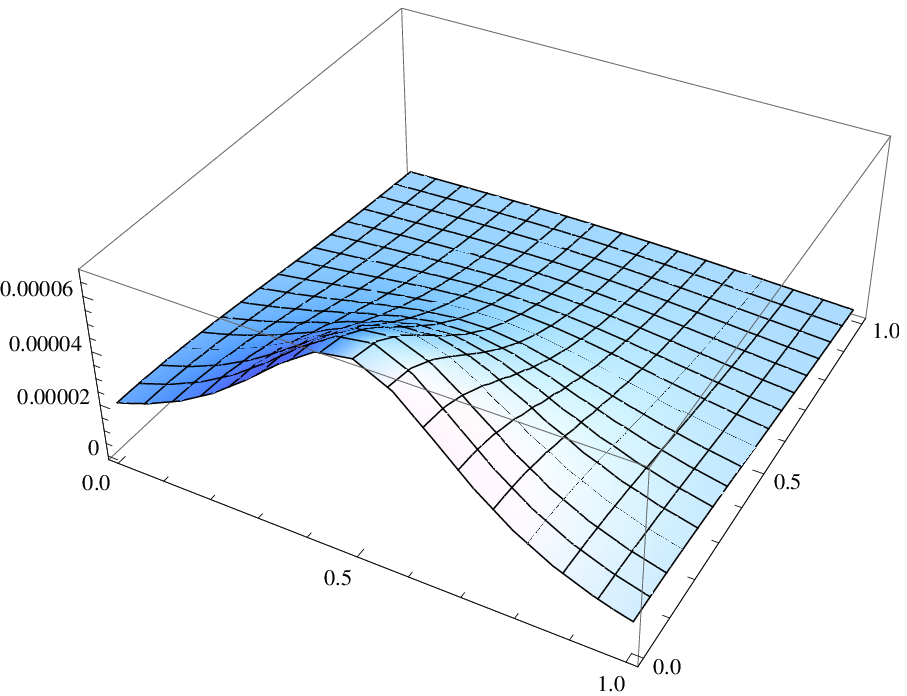}
\caption{\label{bra} Branching ratio of the decay $h\rightarrow c\bar{b} W^-$ for $(\theta_{12},\,\theta_{13})$ bellows to region $[0,1]\times[0,1]$ taking $|\chi|=0.1$, $\delta=0.01$ and $\tan\beta=1$ in scenario \emph{A}.}
\end{minipage}\hspace{3pc}%
\begin{minipage}{14pc}
\includegraphics[width=18pc]{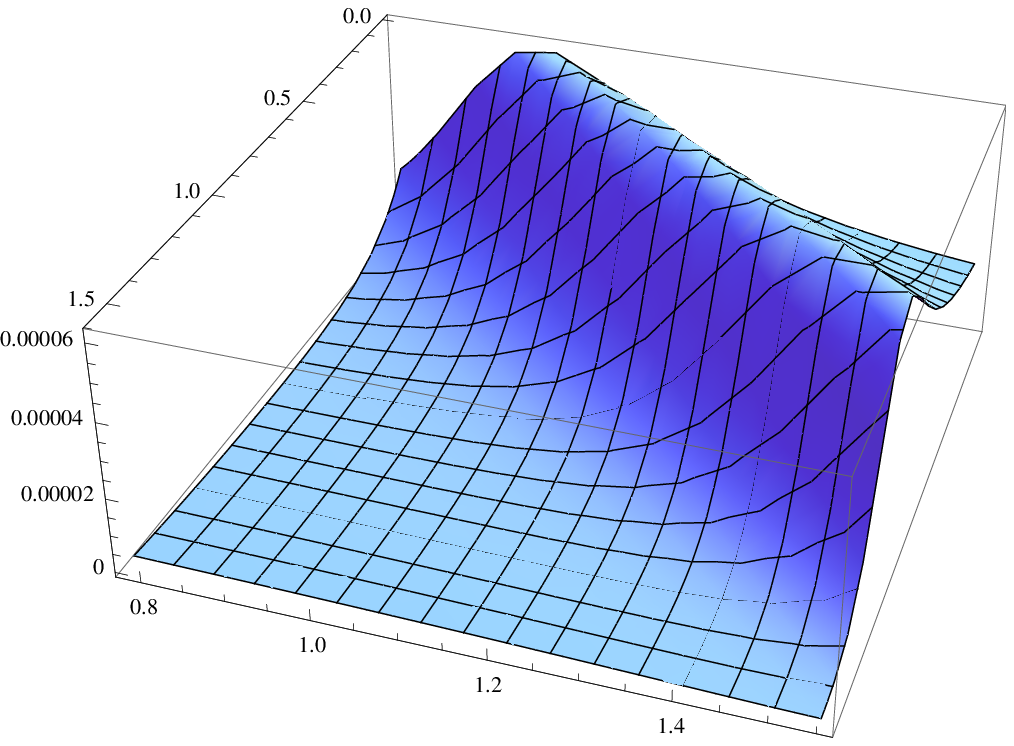}
\caption{\label{brb} Branching ratio of the decay $h\rightarrow c\bar{b} W^-$ for $(\alpha,\,\beta)$ bellows to region $[0,\pi/2]\times[\arctan{1},\arctan{50}]$ taking $|\chi|=0.1$ and $\delta=0.01$ in scenario \emph{B}.}
\end{minipage}
\end{figure}
\begin{figure}
\begin{minipage}{14pc}
\includegraphics[width=18pc]{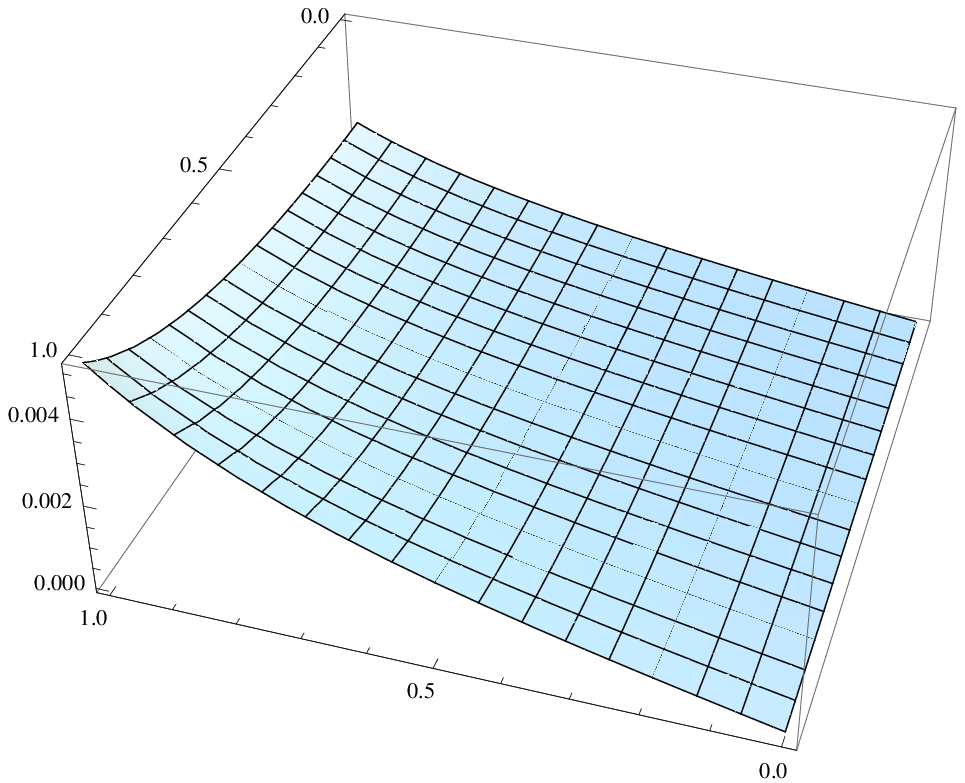}
\caption{\label{acpa} CP asymmetry of the decay $h\rightarrow c\bar{b} W^-$ for $(\theta_{12},\,\theta_{13})$ bellows to region $[0,1]\times[0,1]$ taking $|\chi|=0.1$, $\delta=0.01$ and $\tan\beta=1$ in scenario \emph{A}.}
\end{minipage}\hspace{3pc}%
\begin{minipage}{14pc}
\includegraphics[width=18pc]{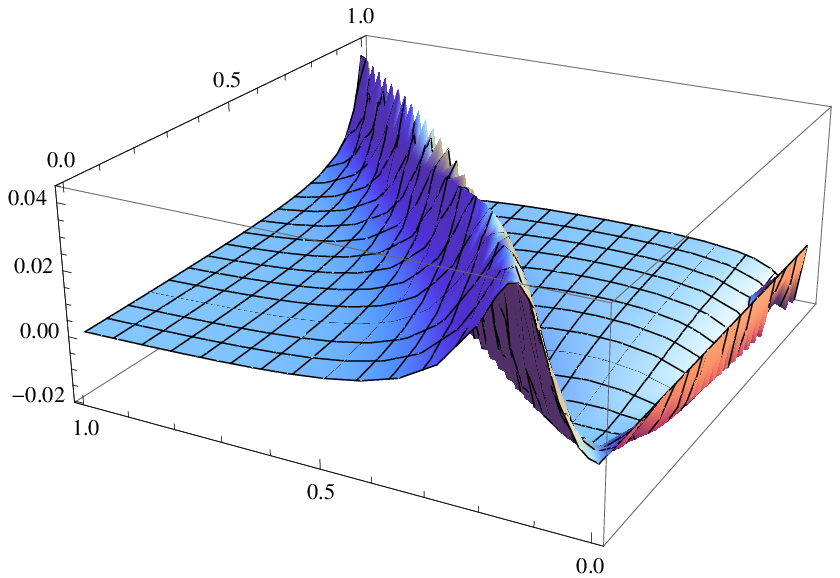}
\caption{\label{acpb} CP asymmetry of the decay $h\rightarrow c\bar{b} W^-$ for $(\alpha,\,\beta)$ bellows to region $[0,\pi/2]\times[\arctan{1},\arctan{50}]$ taking $|\chi|=0.1$ and $\delta=0.01$ in scenario \emph{B}.}
\end{minipage}
\end{figure}
The central numerical values for the SM parameters are the reported by \cite{pdg}. Nevertheless, the mixing angles, texture parameters and neutral scalar mass are free parameters of the 2HDM. However, If the neutral scalar is considered like the Higgs boson scalar in the SM, the current experimental results and data analysis allow to take a central value equal to 125 GeV for the neutral scalar mass \cite{atlas}. Additionally, the ratio of the vacuum expectation values, $\beta$ parameter, is bounded by the region $1\leq \tan\beta \leq 50$ \cite{pdg}.

We have noted that the texture parameters, $\chi_{ij}$, have a small contribution in the branching ratio for both considered scenarios. Then, $\chi_{23}=\chi e^{i\delta}$ are taken as same magnitude order as the CKM parameter, that is, $\chi \approx 0.1 $ and $\delta \approx 0.01$. In figures  and   the dependence of the branching ratio on the mixing angles is shown. The textures parameters controls the magnitude order of the CP asymmetry coefficient defined previously, see (\ref{acpa}), (\ref{acpb}). The CP asymmetry coefficient behavior is shown in figures  and  for the same numerical values of $\chi_{23}$.

\section{Conclusion}
In summary, we have considered a general scalar potential written in $U(2)$ covariant form and the Yukawa couplings of the tipe III with 4-texture ansatz. Under these assumptions the CP violation such as spontaneous or explicit from scalar sector are not the only source but Yukawa couplings contribute. For sake of simplicity, the spontaneous CP violation phase was assumed to be zero and the CP asymmetry coefficient has been obtained. The CP asymmetry coefficient not only depends of mixing angles but texture parameters of the Yukawa couplings. The explicit CP violation and CP conserving possibilities have been considered for scalar sector, which were named as scenario \emph{A} and \emph{B} respectively. The CP asymmetry coefficient has the order of magnitude from $0$ to $10^{-2}$ for both scenarios.

The total width decay in the considered model was approximated by the (\ref{wtotal}). Then an approximated expressions for the branching ratio of the decay $h\rightarrow c\bar{b} W^-$ were obtained for the proposed scenarios. The order of magnitude bellows to range $[0,4\times10^{-5}]$ for $0\leq \theta_{12, 13}\leq \pi/2$  taking scenario \emph{A}, meanwhile the order of magnitude bellows to range $[0,6\times10^{-5}]$ for $0\leq \alpha \leq \pi/2$ and $1\leq \tan\beta\leq 50$  taking scenario \emph{B}.

The decay $h\rightarrow c\bar{b} W^-$ is a important signal for this type of model, 2HDM-III, which could be discriminated from others types of models.
\section*{Acknowledgments}
This work is supported in part by PAPIIT project IN117611-3, Sistema
Nacional de Investigadores (SNI) in M\'exico. J.H.Montes de Oca Y.
is thankful for support from the postdoctoral DGAPA-UNAM grant.
\appendix
\section{The integrals $I_i$}
Defining the dimensionless variables as $\left(
\begin{array}{cc}
x, & y%
\end{array}%
\right) =\left(
\begin{array}{cc}
\frac{2E_{1}}{m_{h}}, & \frac{2E_{2}}{m_{h}}%
\end{array}%
\right) $, $\mu _{1}=\frac{m_{W}^{2}}{m_{h}^{2}}$, $\mu =\frac{m_{t}^{2}}{%
m_{h}^{2}}$, and $\Gamma ^{2}=$ $\frac{\Gamma _{t}^{2}}{m_{h}^{2}}$, the
integrals are
\begin{eqnarray*}
I_{1} &=&\int \int_{R_{xy}}\left( \left( x+y-1-\mu \right) ^{2}+\mu \Gamma
^{2}\right) ^{-1} \\
&&\left[ 4\mu _{1}\left( x+y-\mu _{1}-1\right) \left( x+y+\mu _{1}-1\right)
\left( 2-x-y\right) \right.  \\
&&2\left( x+y-\mu _{1}-1\right) \left( 2-x-y\right)  \\
&&-\left( x+y-1\right) \left( 1-x+\mu _{1}\right)  \\
&&\left. -2\mu _{1}\left( x+y-1\right) \left( x+y-\mu _{1}-1\right) \left(
1-y-\mu _{1}\right) \right] dxdy,
\end{eqnarray*}
\begin{eqnarray*}
I_{2} &=&\int \int_{R_{xy}}\left( \left( x+y-1-\mu \right) ^{2}+\mu \Gamma
^{2}\right) ^{-1} \\
&&\left[ 2\mu \mu _{1}\left( x+y-\mu _{1}-1\right) \left( 1-y-\mu
_{1}\right) \right.  \\
&&\left. \mu \left( 1-x+\mu _{1}\right) \right] dxdy,
\end{eqnarray*}
\begin{eqnarray*}
I_{3} &=&\int \int_{R_{xy}}\left[ \left( x-1\right) ^{-2}\mu _{1}\left(
1-x+\mu _{1}\right) \right.  \\
&&\left. +2\left( x+y-\mu _{1}-1\right) \left( 1-y-\mu _{1}\right) \right]
dxdy,
\end{eqnarray*}%
\begin{eqnarray*}
I_{4} &=&\int \int_{R_{xy}}\left[ \left( \left( x+y-1-\mu \right) ^{2}+\mu
\Gamma ^{2}\right) ^{-1}\left( 1-x\right) ^{-2}\right.  \\
&&\left[ +2\frac{\mu }{\mu _{1}}\left( x+y+\mu _{1}-1\right) \left( 1-y-\mu
_{1}\right) \right.  \\
&&\left. +\mu \mu _{1}\left( 1-x+\mu _{1}\right) \right] \left( x+y-\mu
\right) dxdy,
\end{eqnarray*}%
\begin{eqnarray*}
I_{5} &=&\int \int_{R_{xy}}\left[ \left( \left( x+y-1-\mu \right) ^{2}+\mu
\Gamma ^{2}\right) ^{-1}\left( 1-x\right) ^{-2}\right.  \\
&&\left[ +2\frac{\mu }{\mu _{1}}\left( x+y+\mu _{1}-1\right) \left( 1-y-\mu
_{1}\right) \right.  \\
&&\left. +\mu \mu _{1}\left( 1-x+\mu _{1}\right) \right] \sqrt{\mu }\Gamma
dxdy,
\end{eqnarray*}%
where the $R_{xy}$\ region is defined by
\[
\frac{1}{2}\left( 2-x-\sqrt{x^{2}-4\mu _{1}}\right) \leq y\leq \frac{1}{2}%
\left( 2-x+\sqrt{x^{2}-4\mu _{1}}\right)
\]%
and%
\[
2\sqrt{\mu _{1}}\leq x\leq 1+\mu _{1}.
\]
\begin{figure}
\begin{minipage}{14pc}
\includegraphics[width=18pc]{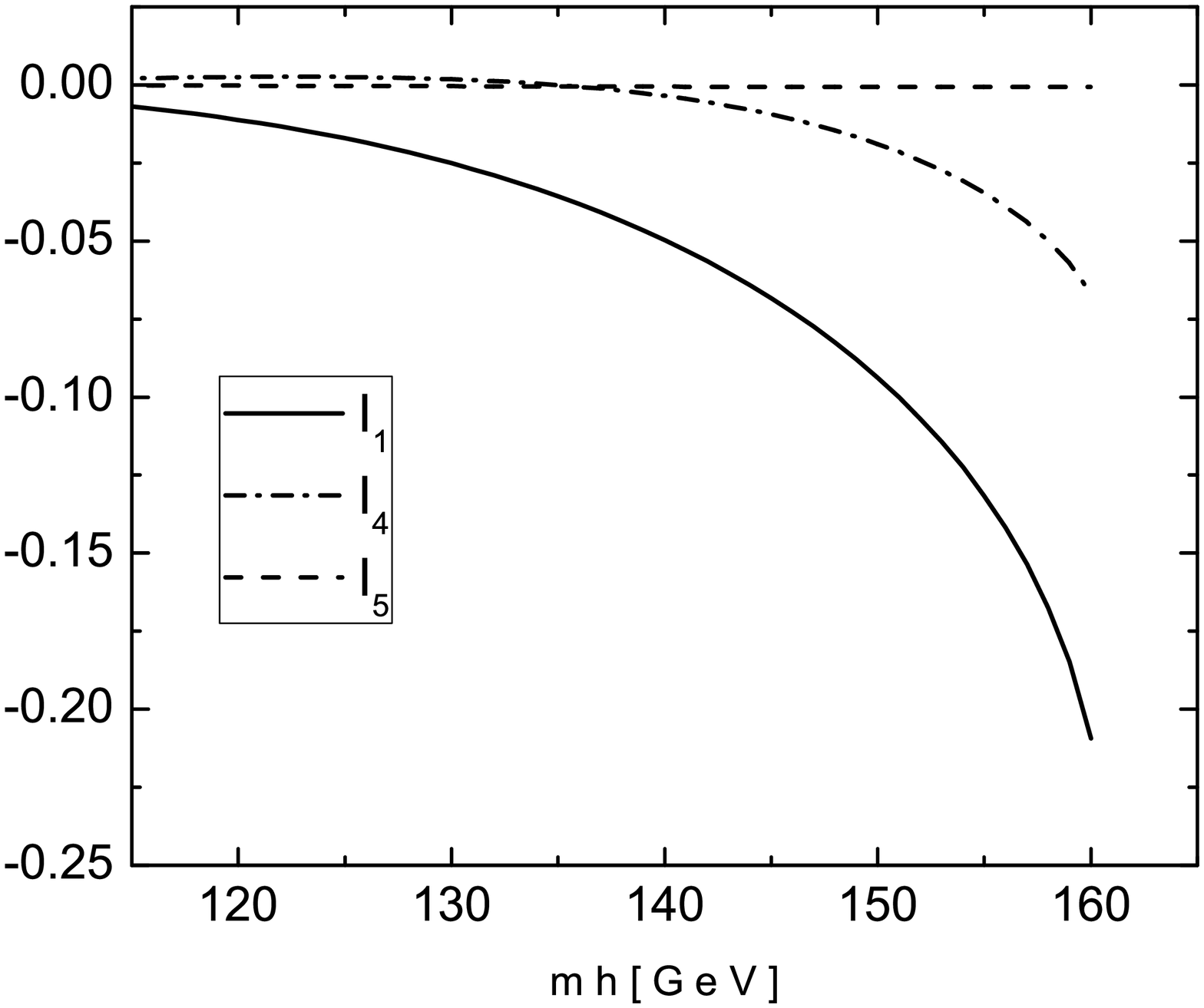}
\caption{\label{i1a} Numerical integrals $I_{1,4,5}$.}
\end{minipage}\hspace{3pc}%
\begin{minipage}{14pc}
\includegraphics[width=18pc]{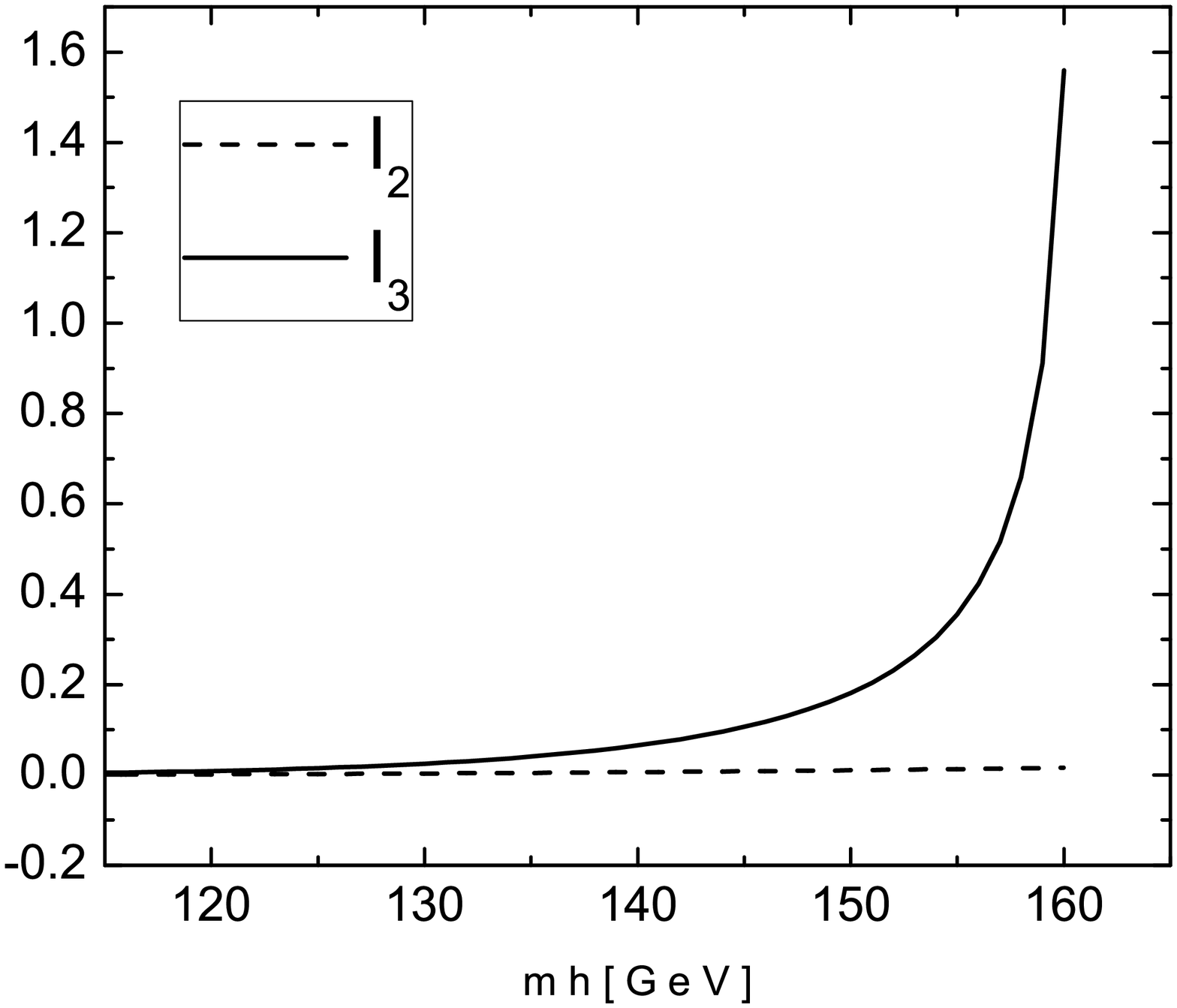}
\caption{\label{i1b} Numerical integrals $I_{2,3}$.}
\end{minipage}
\end{figure}
In figures \ref{i1a} and \ref{i1b} the dependence of the integrals $I_{1,...,5}$ on the scalar mass $h$ is shown.
%

%
%

\end{document}